\documentclass[aps,prl,notitlepage,tightenlines,nofootinbib,twocolumn,superscriptaddress]{revtex4-2}

\usepackage{amsmath}
\usepackage{amssymb,amsthm}
\usepackage{mathtools}
\usepackage{mathrsfs}
\usepackage{relsize}
\usepackage{bm}
\usepackage{ulem}

\usepackage{epsfig,graphics,graphicx,color,xcolor}
\usepackage{soul} 
\usepackage{cancel} 
\usepackage{slashed}	
\usepackage{subfigure}
\usepackage{array}
\usepackage{ragged2e}
\usepackage{lineno}
\usepackage{natbib}
\usepackage{hyperref}
\hypersetup{colorlinks=true,linkcolor=red,anchorcolor=red,citecolor=orange, filecolor=brown,urlcolor=red,bookmarksnumbered=true,
pdfview=FitB
}
\graphicspath{ {image/} }
\usepackage{multirow}
\usepackage{enumitem}

\colorlet{darkgreen}{green!50!black}
\colorlet{darkblue}{blue!70!black}
\colorlet{brightyellow}{yellow!75!red}
\colorlet{orange}{red!50!yellow}
\colorlet{darkgray}{gray!50!black}

\def\fauxschelper#1 #2\relax{%
  \fauxschelphelp#1\relax\relax%
  \if\relax#2\relax\else\ \fauxschelper#2\relax\fi%
}
\def\Hscale{.85}\def\Vscale{.74}\def\Cscale{1.12}
\def\fauxschelphelp#1#2\relax{%
  \ifnum`#1>``\ifnum`#1<`\{\scalebox{\Hscale}[\Vscale]{\uppercase{#1}}\else%
    \scalebox{\Cscale}[1]{#1}\fi\else\scalebox{\Cscale}[1]{#1}\fi%
  \ifx\relax#2\relax\else\fauxschelphelp#2\relax\fi}
  
\def\dd{{\mathrm{d}}}

\newcommand{\half}[1][1] {\mathsmaller{\frac{#1}{2}}}

\makeatletter
\newcommand*{\transpose}{%
  {\mathpalette\@transpose{}}%
}
\newcommand*{\@transpose}[2]{%
  \raisebox{\depth}{$\m@th#1\intercal$}%
}
\makeatother 

\begin{document}

\title{Minkowski's lost legacy and hadron electromagnetism}

\author{Yang~Li}
\affiliation{Department of Modern Physics, University of Science and Technology of China, Hefei 230026, China}

\author{Wen-bo Dong}
\affiliation{Department of Modern Physics, University of Science and Technology of China, Hefei 230026, China}

\author{Yi-liang Yin}
\affiliation{Department of Modern Physics, University of Science and Technology of China, Hefei 230026, China}

\author{Qun Wang}
\affiliation{Department of Modern Physics, University of Science and Technology of China, Hefei 230026, China}

\author{James P. Vary}
\affiliation{Department of Physics and Astronomy, Iowa State University, Ames, IA 50010, U.S.}

\date{\today}

\begin{abstract}
We revisit Minkowski's lost legacy on relativistic electromagnetism in order to resolve long-standing puzzles over the charge distribution of relativistic systems like hadrons. Hadrons are unique relativistic electromagnetic systems characterized by their comparable size and Compton wavelength $r_h \sim \lambda_C$. As such, it was recently realized that the traditional Sachs definition of the charge distribution based on a non-relativistic formula is invalid. We explain that this is the same problem pursued by Lorentz, Einstein and others, on the electromagnetism of a moving body. We show how various charge distributions proposed in hadronic physics naturally emerge as the multipole moment densities in the macroscopic theory of relativistic electromagnetism. 
\end{abstract}
\maketitle

\section{Introduction}

At the turn of the 20th century, some of the greatest minds,  Sir Thomson,  Lorentz and  Einstein along with many others were pursuing -- what now might be regarded as  a ``wrong'' -- problem: the electromagnetic structure of the electron \cite{Minkowski:1908, Einstein:1908}. Later research revealed that the electron is a point-like particle without internal structures \cite{Asbury:1967zzb}. Nevertheless, relativity, a pillar of modern physics, was born from these endeavors. The relativistic theory of macroscopic electromagnetism was fully established by Minkowski, Einstein and Laub in 1908 \cite{Pauli:1981}. 
The final expressions Einstein, Minkowski and Laub derived for moving media are identical to the macroscopic Maxwell equations applicable to media at rest. 
Therefore, it is then not a surprise when the right kind of problem finally arrived in the 1950s, the investigation of the proton electromagnetic structure \cite{Yennie:1957, Hofstadter:1958}, their work was largely forgotten. The proton charge distribution was then defined based on the non-relativistic formula \cite{Ernst:1960zza, Sachs:1962zzc}, 
 \begin{equation}\label{eqn:Sachs_FF}
\rho_\text{ch}^{(\text{Sachs})}(\vec x) \equiv \int \frac{\dd^3 q}{(2\pi)^3} \, e^{-i\vec q\cdot \vec x} G_E(-\vec q^2).
\end{equation}
In this formula, proposed by Sachs et al., $G_E(q^2)$ is the charge form factor obtained from the covariant decomposition of the hadron matrix element, $\langle p' | J^\mu(0)|p\rangle$, which can be measured in elastic electron-proton scatterings. This expression is usually understood as defined in the Breit frame $q^0 = 0$, $\vec p + \vec p' = 0$. 
It is known to be ambiguous at large $Q^2 = |q^2|$. For example, the Dirac form factor $F_1$ is an equally good choice as the charge form factor. It is probably an ad hoc expression, since at that time, the accessible $Q^2$ was quite limited.  Nevertheless, this has become the textbook definition of the charge distribution \cite{Kelly:2002if, Perdrisat:2006hj, Pacetti:2014jai, Wong:1998ex, Suhonen:2007vjh}. 

Only recently, the Sachs charge distribution was criticized as unphysical \cite{Miller:2007uy, Miller:2009sg, Miller:2009qu, Miller:2018ybm, Jaffe:2020ebz, Freese:2021czn, Freese:2021mzg}. Miller et al. argued that a physically measurable charge distribution should be obtained from the quantum expectation value $\rho_\Psi(x) = \langle\Psi|J^0(x)|\Psi\rangle$, 
which depends on the hadron wavepacket $\Psi(\vec p) = \langle p | \Psi\rangle$. In relativistic quantum mechanics (including quantum field theories), the Lorentz boosts are dynamical \cite{Dirac:1949cp}. As a result, the intrinsic charge distribution of the hadron cannot be separated from the c.m. motion. For example, with a Gaussian wavepacket of a width $R_\text{w}$, the r.m.s. charge radius $\langle r^2_\text{ch} \rangle_{\Psi} = -6G'_E(0) + 3R^2_\text{w}$ diverges as the wavepacket  approaches the plane wave $R_\text{w} \to \infty$, which is the implicit wavepacket adopted in the Sachs definition \cite{Freese:2021czn}. 
Similarly, a localized wavepacket is not possible, either, since particles in relativistic quantum mechanics cannot be localized \cite{localization, Newton:1949cq, Wightman:1962sk, Haag:1992hx, Busch:1999, Balachandran:2016bqj}.  
One could give up the 3D description in favor of a 2D light-front distributions \cite{Burkardt:2000za} with all its peculiarities \cite{Miller:2007uy}. Indeed, the majority of the works in hadron structure are devoted to this formulation \cite{Burkardt:2002hr}. 
To resolve these issues, Lorcé applied the Weyl-Wigner quasi-distributions defined in a special reference frame, the elastic frame $q^0 = 0$. The Sachs distribution and the light-front distribution appear as two slices of the quasi-distributions \cite{Lorce:2020onh}. Epelbaum et al. provided a definition based on sharply localized wavepackets with a spherical symmetry \cite{Epelbaum:2022fjc}.

Jaffe further clarified that, in order for the Sachs definition to be valid, a chain of inequalities have to be satisfied \cite{Jaffe:2020ebz} (cf. Refs.~\cite{Belitsky:2003nz, Ji:2004gf, Belitsky:2005qn})
\begin{equation}
r_h \gg \lambda_\gamma \gg \lambda_h \ge \lambda_\text{C},
\end{equation}
where, $r_h$ is the hadron radius, $\lambda_\gamma \sim Q^{-1}$ is the wavelength of the probing photon; $\lambda_h$ is the de Broglie wavelength of the hadron, which is bounded by its Compton wavelength $\lambda_\text{C} = M^{-1}_h$. Here, 
$M_h$ is the hadron mass and we have adopted the natural units $\hbar = c = 1$. Unfortunately, the above inequalities do not hold, since for relativistic systems like hadrons, their size and their Compton wavelength are comparable $r_h \sim \lambda_\text{C}$.  For example, the proton charge radius is $r_p \simeq 0.84\,\text{fm}$, while its Compton wavelength is $M^{-1}_p \simeq 0.2\,\text{fm}$ \cite{PDG:2020}. The mass radius of the proton is even smaller $r_p^{\text{mass}} \simeq 0.55\,\text{fm}$ \cite{Polyakov:2018zvc, Kharzeev:2021qkd}. Similarly, the pion charge radius is $r_\pi \simeq 0.67\,\text{fm}$ whereas its Compton wavelength is $M^{-1}_\pi\simeq 1.4\,\text{fm}$ \cite{PDG:2020}.

An immediate implication of Jaffe's argument is that a proton is a relativistic matter wave, as far as the electromagnetic probe ($\lambda_\gamma \lesssim \lambda_h$) is concerned. This view takes us back to Einstein and Minkowski's problem, now with the right kind of system to investigate. Minkowski, Einstein and Laub's theory of relativistic electromagnetism tells us what kind of quantities can be extracted from the measurements. For example, the matter wave as a medium is characterized by an antisymmetric medium polarization tensor $M^{\alpha\beta}$ whose components give the polarization vector $\vec{{P}}^i = M^{0i}$ and the magnetization vector $\vec{{M}}^i = -\frac{1}{2}\epsilon^{ijk}M^{jk}$. The full current is the sum of the free current -- the current of a poin-like source -- and the polarized current, $j^\beta = j^\beta_\text{f} + \partial_\alpha M^{\alpha\beta}$. The coupling to the classical electromagnetic field is determined by the Minkowski-Maxwell equations $\partial_\alpha H^{\alpha\beta} = j^\beta_\text{f}$, where the field induction tensor $H^{\alpha\beta} = F^{\alpha\beta} - M^{\alpha\beta}$. 

The macroscopic charge and current densities are associated with these quantities, e.g. the polarization charge density 
$\rho_\text{pol} = -\nabla\cdot\vec{{P}}$, the effective magnetic charge density $\rho_\text{mag} = -\nabla\cdot\vec{{M}}$, and the magnetization current $\vec j_\text{mag} = \nabla\times\vec{{M}}$. One interesting consequence of this formalism is that a magnetized but unpolarized hadron, e.g. the neutron, will acquire an electric dipole moment under relativistic motion \cite{Pauli:1981, Miller:2007uy, Lorce:2020onh}. For media in motion, it is useful to introduce the co-moving polarization 4-vector and the co-moving magnetization 4-vector: $M^{\alpha\beta} =  u^\alpha \mathcal P^\beta  - u^\beta\mathcal P^\alpha  + \varepsilon^{\alpha\beta\rho\sigma}u_\rho \mathcal M_\sigma$, and the associated macroscopic densities, 
$\varrho_\text{pol}(x) = -\partial_\alpha{\mathcal{P}^\alpha}$, $\varrho_\text{mag}(x) = -\partial_\alpha{\mathcal{M}^\alpha}$, where $u^\alpha$ is a timelike unit frame vector, whose definition, however, is not unique \cite{Israel:1979wp, Eckart:1940te, Landau:vol6}.

For electromagnetic media comprised of composite particles, the multipole moment densities describe the internal electromagnetic structures of these composite particles \cite{Jackson:1999, deGroot:1972}. We will see that these multipole moment densities bridge the macroscopic observables and the conventional quantities like the Sachs distribution. On the other hand, they also lead to new characterizations of the system. 
It is worth mentioning that what we discuss in this work has nothing to do with the proton radius puzzle \cite{Pohl:2013yb, Bezginov:2019mdi, Xiong:2019umf}. 

While we focus on the electromagnetic structures here, the formulation can be readily generalized to hadron structures under the weak, gravitational and Higgs probes, e.g.~\cite{Selyugin:2009ic, Polyakov:2018zvc}. In particular, the gravitational form factors have been analyzed using similar frameworks as we proposed here \cite{ Freese:2022fat, Panteleeva:2022uii, Alharazin:2022xvp}.

\section{Densities in quantum field theory}

Consider a relativistic quantum system, whose internal dynamics is described by an underlying relativistic quantum field theory. For simplicity, we start with a spin-0 hadron.  
The full current is given by, 
\begin{multline}
j^\mu(x) = \langle \Phi | J^\mu(x) | \Phi\rangle 
 = \int \frac{\dd^3 p}{(2\pi)^32p^0} \int \frac{\dd^3 p'}{(2\pi)^32p'^0} \\
 \times \widetilde\Phi^*(\vec p')\widetilde\Phi(\vec p) 2 P^\mu F_\text{ch}(q^2) e^{iq\cdot x}, \label{eqn:full}
\end{multline}
where $p$ and $p'$ are on-shell momenta with  $p^0 = \sqrt{\vec p^2+M^2}$, $q = p'-p$ and $P = \frac{1}{2}(p'+p)$. The charge form factor $F_\text{ch}(q^2)$ is defined from the hadron matrix element of the current operator $J^\mu$, viz. 
$\langle p' | J^\mu(0) | p \rangle = (p+p')^\mu F_\text{ch}(q^2)$.
The hadron state vector is normalized as $\langle\Phi|\Phi\rangle = 1$ and $\langle p'|p\rangle = 2p^0(2\pi)^3\delta^3(p-p')$. As such, the wavepacket $\widetilde\Phi(\vec p) = \langle p |\Phi\rangle$ is also normalized as,
\begin{equation}
\int \frac{\dd^3 p}{(2\pi)^32p^0} \widetilde\Phi^*(\vec p)\widetilde\Phi(\vec p) = 1.
\end{equation}
We can introduce a \textit{Lorentz covariant} ``coordinate space wave function'' as the Fourier transform of the momentum space wave function, 
\begin{equation}\label{eqn:coord}
\Phi(x) = \int \frac{\dd^3 p}{(2\pi)^32p^0} \widetilde\Phi(\vec p) e^{-i p\cdot x},
\end{equation}
which satisfies the Klein-Gordon (KG) equation. In contrast to their momentum space counterpart, the KG wave function cannot be normalized, since it is not a true wave function. Instead, it is normalized as a current, 
\begin{equation}
\int \dd^3 x\, \Phi^*(x)i\tensor{\partial}_t \Phi(x) = 1,
\end{equation}
where, $f\tensor{\partial} g \equiv  f \partial g - (\partial f)g$. This fact is closely related to the non-existence of the coordinate operator in relativistic quantum theory. Using the KG wave function, we define the free current as the convective current of a point-like particle \cite{Weinberg:1995mt}, 
$j^\mu_\text{f}(x) =  \Phi^*(x) i\tensor{\partial}^\mu \Phi(x)$.
The medium polarization tensor can be constructed as,
\begin{multline}\label{eqn:medium_polarization}
M^{\mu\nu}(x) 
 = \int \frac{\dd^3  P}{(2\pi)^3} \int \frac{\dd^3 q}{(2\pi)^3} \widetilde\Phi^*({\vec P}+\half\vec q)\widetilde\Phi({\vec P}-\half\vec q) \\
 \times \frac{ q^\mu P^\nu - q^\nu P^\mu}{2p^0p'^0} \frac{F_\text{ch}(q^2)-1}{i q^2} e^{iq\cdot x}.
 \end{multline}
 From this expression, we obtain the macroscopic charge and current densities $\rho_\text{pol}$, $\rho_\text{mag}$ and $\vec j_\text{mag}$. For example, 
  \begin{multline}
 \rho_\text{pol}(x) = \int \frac{\dd^3 P}{(2\pi)^3} \int \frac{\dd^3 q}{(2\pi)^3} \widetilde\Phi^*({\vec P}+\half\vec q)\widetilde\Phi({\vec P}-\half\vec q)\\
 \times \frac{P^0}{2p^0p'^0} \big(F_\text{ch}(q^2)-1\big) e^{iq\cdot x}.
\end{multline}
For the co-moving densities, we need to define the frame vector $u^\alpha$ first. Different schemes exist. A very simple choice is $u^\alpha = i\tensor{\partial}^\alpha/\sqrt{-\tensor{\partial}\cdot\tensor{\partial}} = 2P^\alpha/\sqrt{4M^2 - {q^2}}$. The corresponding 
polarization charge density is,
\begin{multline}
\varrho_\text{pol}(x)  
 = \int \frac{\dd^3 p}{(2\pi)^32p^0} \widetilde\Phi(\vec p)\int \frac{\dd^3 p'}{(2\pi)^32p'^0}  \widetilde\Phi^*(\vec p') \\
 \times\sqrt{4M^2-q^2} \big(F_\text{ch}(q^2)-1\big) e^{iq\cdot x}.
\end{multline}

Similarly, for a spin-1/2 particle, the full current is,
\begin{multline}
j^\mu(x) =  \int\frac{\dd^3p}{(2\pi)^32p^0}\int\frac{\dd^3 p'}{(2\pi)^32p'^0}
\widetilde\Psi_{s'}^*(\vec p') \widetilde\Psi_{s}(\vec p) \\
\times \bar u_{s'}(p')\Big[\gamma^\mu F_1(q^2) + \frac{i\sigma^{\mu\nu}q_\nu}{2M}F_2(q^2)\Big] u_s(p) e^{iq\cdot x},
\end{multline}
where, $F_1, F_2$ are the Dirac and Pauli form factors, respectively. The free current is obtained from taking the point particle limit,
\begin{equation}
j_\text{f}^\mu(x) = \frac{q_N}{2M}\overline\Psi(x) i\tensor{\partial}^\mu \Psi(x) + \mu_N\partial_\nu \big[ \overline\Psi(x) \sigma^{\mu\nu} \Psi(x)\big],
\end{equation}
where, $M$ is the spinor mass, $q_N=F_1(0)$ is its charge number and $\mu_N=[F_1(0) + F_2(0)]/2M$ is its magnetic moment. 
The Lorentz covariant ``wave function'' 
\begin{equation}
\Psi(x) = \sum_s \int \frac{\dd^3 p}{(2\pi)^32p^0} \widetilde\Psi_s(\vec p) u_s(p) e^{-i p\cdot x}
\end{equation}
satisfies the Dirac equation. 
The medium polarization tensor can be constructed as, 
\begin{multline}\label{eqn:polarization_tensor_spoinor}
M^{\mu\nu} =
\int \dd^3 r_1 \int \dd^3 r_2 \,
 \overline \Psi(\vec r_1, t) 
 \int \frac{\dd^3 P}{(2\pi)^3}  e^{i{\vec P}\cdot(\vec r_1-\vec r_2)}  \\
 \times \int \frac{\dd^3 q}{(2\pi)^3} \Big\{(q^\mu \gamma^\nu - q^\nu \gamma^\mu) \frac{F_1(q^2) - F_1(0)}{iq^2} \\
 - \frac{\sigma^{\mu\nu}}{2M} \big[F_2(q^2) - F_2(0)\big]  \Big\}  
 e^{- i\vec q\cdot (\vec x-\frac{\vec r_1+\vec r_2}{2})}
 \Psi(\vec r_2, t). 
\end{multline}

\begin{figure}
\centering
\includegraphics[width=0.4\textwidth]{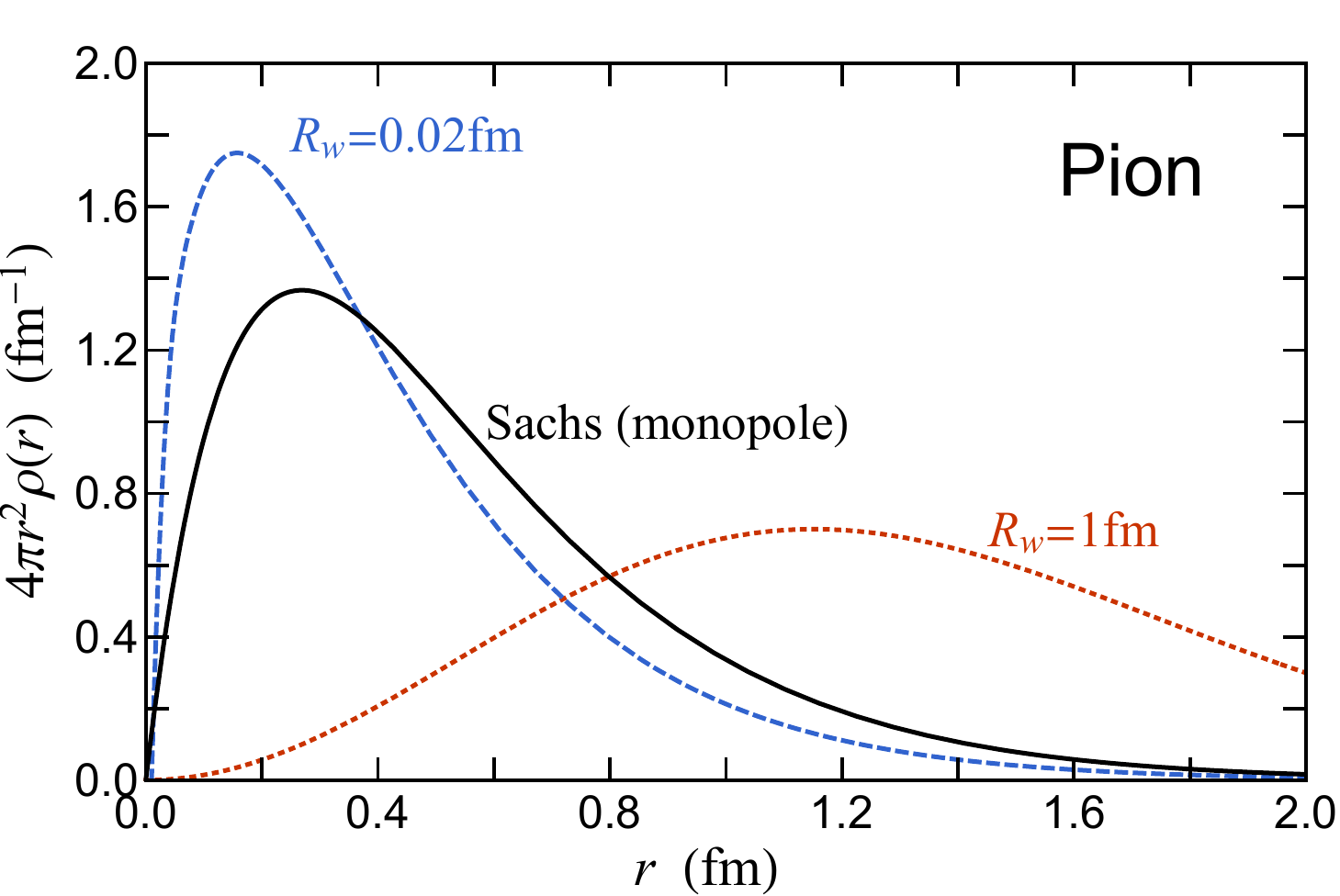}
\caption{Charge distribution $\rho(r) = j^0(r)$ with Gaussian wavepackets $\widetilde\Phi(\vec p) = N \exp(-\frac{1}{2} R_w^2\vec p^2 )$, as compared with the Sachs distribution. We adopt a dipole ansatz for the pion form factor. The dependence on the wavepacket is shown. For the pion, the charge distribution in neither the localization limit ($R_w \to 0$) or the plane wave limit ($R_w \to \infty$)  reproduces the Sachs distribution which, as we explain later in the paper, is in fact the electric monopole density.}
\label{fig:pol}
\end{figure}

Figure~\ref{fig:pol} shows the charge density of the pion with selected Gaussian wavepackets as compared with the 
Sachs distribution (\ref{eqn:Sachs_FF}). Neither the localization limit ($R_w\to 0$) or the plane wave limit ($R_w\to\infty$) 
reproduces the Sachs distribution. This comparison may reinforce the traditional impression that the Sachs distribution 
is only an approximation valid in the nonrelativistic limit. However, as we will see later, the Sachs distribution is in fact an exact 
quantity, the electric monopole density. 

\section{Hadronic  multipole moment densities}

\begin{figure}
\centering
\includegraphics[width=0.3\textwidth]{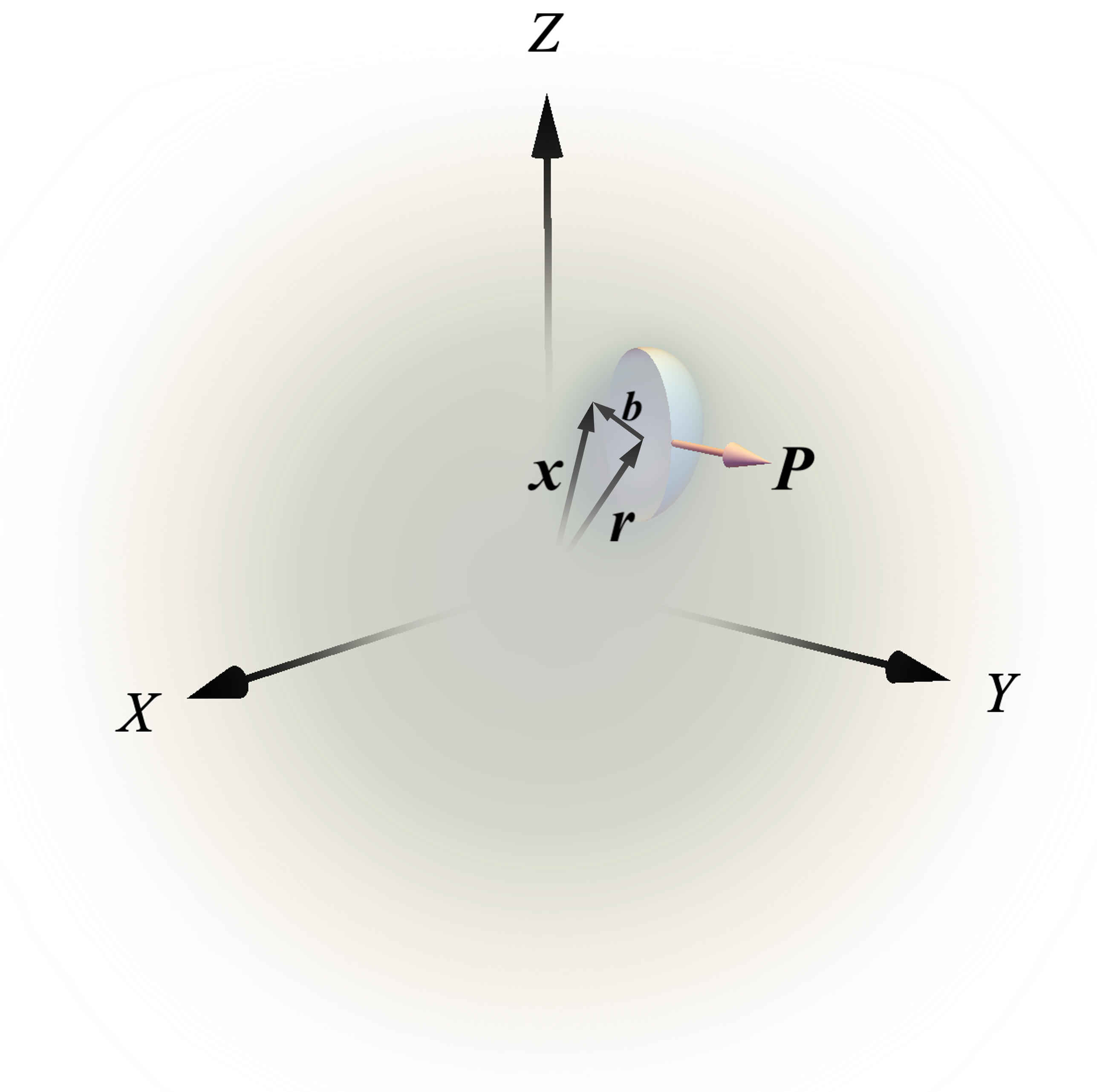}
\caption{The full current can be viewed as the convolution of the hadronic multipole density and the wavepacket. $\vec r$ and $\vec P$ are the coordinate 
and the momentum of the hadron (shown as a Lorentz contracted ellipsoid), respectively. $\vec b$ is the coordinate of the internal charge, conjugate to $\vec q$.}
\label{fig:atomic_moments}
\end{figure}

The macroscopic densities depend on the wavepacket of the hadron. A central goal in hadronic physics is to extract the intrinsic information of the subatomic particles in 3D. 
In Minkowski's relativistic electromagnetism, the internal distribution of the composite particles are characterized by the multipole moment densities \cite{deGroot:1972}. Schematically, the classical current may be written as a convolution of the convective current and the multipole density $\varTheta(\vec r)$ (see Fig.~\ref{fig:atomic_moments}),
\begin{equation}\label{eqn:convolution}
j^\mu(x) \overset{\cdot}{=} \int \dd^3r\, \varTheta (\vec x - \vec r) j^\mu_\text{f}(\vec r, t).
\end{equation}
From Eqs.~(\ref{eqn:full}--\ref{eqn:coord}), we obtain, 
\begin{multline}\label{eqn:full_current}
j^\mu(x) 
 = \int\dd^3r_1 \Phi^*(\vec r_1, t)  \int \dd^3 r_2 i \partial^\mu \Phi(\vec r_2, t) 
  \int \frac{\dd^3 P}{(2\pi)^3} \\
  \times e^{i\vec P\cdot (\vec r_1-\vec r_2)} 
 \int \frac{\dd^3 q}{(2\pi)^3} F_\text{ch}(q^2) e^{i\vec q\cdot (\frac{\vec r_1+\vec r_2}{2}-\vec x)} 
 + \text{H.c.}
\end{multline}
Recall, $q^2 = (q^0)^2 - \vec q^2$, which depends on $\vec P$ as,
\begin{equation}
q^0 = \sqrt{(\vec P+\half\vec q)^2+M^2} - \sqrt{(\vec P-\half\vec q)^2+M^2}.
\end{equation}
Comparing to the classical case (\ref{eqn:convolution}),  we need to evaluate the $\int\dd^3P$ integral in Eq.~(\ref{eqn:full_current}) to obtain a Dirac-$\delta$, provided $\vec P$ in the remainder of the expressions is
replaced by $(-i/2)\tensor{\nabla}_{\vec r}$. The resulting expression is,
\begin{equation}
j^\mu(x) 
 = \int\dd^3r\,   \Phi^*(\vec r, t)  \varTheta(\vec x - \vec r) i \partial^\mu \Phi(\vec r, t) 
+ \text{H.c.}
\end{equation}
where, 
\begin{equation}\label{eqn:atomic_moment}
\varTheta(\vec b) = \int \frac{\dd^3 q}{(2\pi)^3} F_\text{ch}(q^2) e^{-i\vec q\cdot \vec b}. 
\end{equation}
Again, it depends on $\vec P=(-{i}/{2})\tensor{\nabla}_{\vec r}$ via $q^0$. This dependence has to be 
defined through a power series. For example, the Taylor series of $F_\text{ch}(q^2)$ around $|\vec P|=0$ is, 
\begin{align}\label{eqn:P0_series}
q^2 =\,&  -\vec q^2 + \frac{(\vec q \cdot \vec P)^2}{M^2+\vec q^2/4} + \mathcal O(|\vec P|^4), \\
F_\text{ch}(q^2) =\,&  F_\text{ch}(-\vec q^2) + \frac{F'_\text{ch}(-\vec q^2)}{M^2+\vec q^2/4} (\vec q \cdot \vec P)^2 
+ \mathcal O(|\vec P|^4). \nonumber
\end{align}
Then, the multipole densities $\mathfrak m^{i_1i_2\cdots i_n}_n$ are defined as, 
\begin{equation}
\varTheta(\vec b) = \sum_{n} \frac{(-i)^n}{2^n n!} \mathfrak m^{i_1i_2\cdots i_n}_n(\vec b) \tensor{\nabla}^{i_1}\tensor {\nabla}^{i_2}\tensor{\nabla}^{i_3}\cdots \tensor{\nabla}^{i_n}.
\end{equation}
With (\ref{eqn:P0_series}), the first two multipole densities are, 
\begin{align}
\mathfrak m_0(\vec r) = \,& \int \frac{\dd^3 q}{(2\pi)^3} F_\text{ch}(-\vec q^2)
e^{- i\vec q\cdot \vec r}, \label{eqn:monopole} \\
\mathfrak m_2^{ij}(\vec r) = \,& 2\int \frac{\dd^3 q}{(2\pi)^3} 
\frac{F'_\text{ch}(-\vec q^2)}{M^2+\vec q^2/4}q^iq^j
e^{- i\vec q\cdot \vec r}.
\end{align}
The monopole density (\ref{eqn:monopole}) is just the Sachs charge distribution (\ref{eqn:Sachs_FF}) -- no special frame is chosen here. Moreover, in addition to the monopole moment, the scalar hadron also possesses higher moments -- an effect of purely relativistic origin: the hadron is Lorentz contracted in the direction of motion within the wavepacket (see Fig.~\ref{fig:atomic_moments}). 

Note that the series expansion is not unique. 
A closely related issue is the convergence of the series. 
Since both the wavepacket and the hadron density span over the entire space, there will always be regions outside the radius of absolute convergence. 
The hope is that the convoluted density is sufficiently suppressed in those regions. 
Roughly speaking, $|\vec P| $ corresponds to the inverse of the de Broglie wavelength, $|\vec P| \sim \lambda^{-1}_h$ and similarly, $|\vec q| \sim \lambda^{-1}_\gamma$. Therefore, the series in (\ref{eqn:P0_series}) converge if the width of the wavepacket is sufficiently large, approaching a plane wave. 
Similarly, convergence of the series in terms of $1/|\vec q|$ requires a localized intrinsic charge density -- not a viable choice for hadrons since $r_\text{ch} \sim M^{-1}$. 

An interesting alternative is the Taylor expansion in terms of $1/|\vec P|$, which requires a wavepacket sufficiently localized in at least one spatial direction. In this case, 
\begin{equation}
q^2 = - \vec q_\perp^2 - \frac{q_\|^2 (M^2+\frac{1}{4}\vec q^2_\perp)}{\vec P^2} + \mathcal O(|\vec P|^{-4}), 
\end{equation} 
where, $q_\| = {\vec q \cdot \vec P}/{|\vec P|}$, and $\vec q_\perp = \vec q -  ({\vec q\cdot \vec P}/{\vec P^2})  \vec P$. 
The resulting monopole density is the light-front distribution, 
\begin{equation}
\mathfrak m_0(\vec r) =  \delta(r_\|) \int \frac{\dd^2 q_\perp}{(2\pi)^2} F_\text{ch}(-\vec q^2_\perp)
e^{- i\vec q_\perp\cdot \vec r_\perp}.
\end{equation}
We emphasize that the convergence of this multipole expansion does not require a fully localized (i.e. localization in 3D) hadron, which is physically problematic in relativistic quantum field theory (cf. \cite{Fleming:1974af, Epelbaum:2022fjc, Panteleeva:2022khw}).

For spinors the multipole densities are not uniquely defined since they are not Lorentz scalars and electric and magnetic quantities may be converted into each other. 
This point is reflected in various choices of the electric and magnetic form factors. 
One of the conventional choices is to associate the electric part with the point-like free Dirac particle, i.e. $\overline\psi \gamma^\mu \psi$ 
and to associate the magnetic part with the spin $\partial_\nu (\overline\psi \sigma^{\mu\nu} \psi)$, 
\begin{multline}
j^\mu(x) 
 = \int\dd^3r\,  \overline\Psi(\vec r, t)  \varTheta_1(\vec x - \vec r) \gamma^\mu \Psi(\vec r, t) \\
+ \int\dd^3r\,  \overline\Psi(\vec r, t) \frac{\sigma^{\mu\nu}}{2M}\partial_\nu \varTheta_2(\vec x - \vec r) \Psi(\vec r, t).\end{multline}
Here, $\varTheta_{1,2}$ are the multipole densities associated with the Dirac and Pauli form factors $F_{1,2}$, respectively.

Alternative partition of the electricity and magnetism is based on the convection 
$\overline\Psi(x) i\tensor{\partial}^\mu \Psi(x)$ and non-convection (i.e. spin) parts of the current \cite{Lorce:2020onh}, which leads to the following multipole expansion, 
\begin{multline}
j^\mu(x) 
 = \frac{1}{2M}\int\dd^3r\,  \overline\Psi(\vec r, t)  \varTheta_E(\vec x - \vec r) i \partial^\mu \Psi(\vec r, t) \\
+ \frac{i\varepsilon^{\mu\nu\alpha\beta} }{2M}\int\dd^3r\,  \overline\Psi(\vec r, t)\gamma_\beta\gamma_5 \Sigma_\nu(\vec x - \vec r) i \partial_\alpha\Psi(\vec r, t) \\
+ \text{H.c.}
\end{multline}
where, $\varTheta_E$ and $\Sigma^\mu = i\partial^\mu \varTheta_M$ are the multipole (current) densities associated with the Sachs form factors $G_E$ and $G_M$, respectively.
In this scheme, magnetic effects induced by charged particle motion should vanish in its local rest frame. 
These different choices are in parallel to the various choices of the fluid frames in spin relativistic hydrodynamics \cite{Israel:1979wp, Florkowski:2018fap, Sheng:2022ssd, Sheng:2022wsy, Sheng:2022ffb}, notably Landau-Lifshitz's energy frame \cite{Landau:vol6} and Eckart's particle frame
\cite{Eckart:1940te}.

\section{Summary and Discussions}

In this work, we revisited the macroscopic field theory description of the electromagnetic structures of hadrons (e.g. proton, pion). This description emerges because of the relativistic quantum nature of hadrons: $r_h \sim \lambda_C$. The macroscopic fields are obtained from the quantum average while  the underlying dynamics is still dictated by quantum field theory. Thus, they are physically measurable quantities. 

The field description brings a salient yet fundamental viewpoint in the quest of hadron structures in 3D.  Minkowski, Einstein and Laub's theory bridges the macroscopic densities with the microscopic observables. The Sachs charge distribution and the light-front charge distributions are clarified as two types of the multipole moment expansion. These quantities can be measured from any frame  -- no special frames are needed -- provided the convergence conditions of the corresponding multipole expansions are met. Of course, the inverse problem of Eq.~(\ref{eqn:convolution}),
known as deconvolution in signal processing, is highly non-trivial even for Gaussian wavepackets \cite{Wiener:1964}. In this regard, the light-front formalism, exploiting (partially) localized wavepackets,  has a clear advantage in accessing the intrinsic structures of hadrons.

As we have mentioned, hadrons are distorted in phase space due to the Lorentz contraction. This effect is dynamical. 
However, one may introduce a rotationless Lorentz boost $\Omega$ with $\Omega  \cdot P = (M, \vec 0)$ to 
define ``intrinsic" multipole densities $\vartheta(\vec r) = \varTheta({\vec \Omega}\cdot\vec r)$ in a momentary rest frame.  As such, the kinematical part of the Lorentz contraction may be removed. However, the choice of $\Omega$ is not unique \cite{Polyzou:2012ut}, leading to different intrinsic moments. One can show that with the canonical boost, only the monopole density, i.e. the Sachs charge distribution, survives  for scalar particles in Eq.~(\ref{eqn:atomic_moment}). N.B. both the spin and the current components will change under the Lorentz boost. Indeed, Rinehimer and Miller showed that boosting 
the current to the infinite momentum frame converts the Sachs form factor $G_E$ to the Pauli form factor $F_1$ \cite{Rinehimer:2009yv}. 

An immediate generalization of the present formalism is to the energy-momentum tensor (EMT), which is the conserved Noether current of the diffeomorphism invariance \cite{Freese:2021jqs}. Using the same techniques, the full EMT current can be written as
two parts as well, $t^{\alpha\beta} \equiv \langle \Phi |T^{\mu\nu}(x)| \Phi \rangle = t^{\alpha\beta}_\text{f} + \partial_\sigma \chi^{\mu\nu\sigma}$, where the free EMT,
\begin{multline}
t^{\alpha\beta}_\text{f} = \partial^\alpha \Phi^* \partial^\beta \Phi + \partial^\beta\Phi^* \partial^\alpha \Phi 
 + D g^{\alpha\beta} \big(\partial_\sigma\Phi^* \partial^\sigma \Phi \\
 -M^2\big|\Phi \big|^2\big) 
 - \frac{1}{2}(D+1)\partial^\alpha\partial^\beta \big|\Phi \big|^2, \label{eqn:free_EMT}
\end{multline}
and the polarization tensor $\chi^{\alpha\beta\sigma}$,
\begin{multline}
\chi^{\alpha\beta\sigma}(x) 
= \int \frac{\dd^3p}{(2\pi)^32p^0}\int \frac{\dd^3p'}{(2\pi)^32p'^0}
\widetilde\Phi^*(\vec p')\widetilde\Phi(\vec p) \\
\times \Big\{
2P^\alpha (P^\beta q^\sigma -  P^\sigma q^\beta) \frac{A(q^2)-1}{iq^2} \\
-2i \big(g^{\alpha\beta}q^\sigma - g^{\alpha\sigma}q^\beta\big) (D(q^2)-D) \Big\}.
\end{multline} 
Here, $D = D(0)$ is a less known global charge, the druck term (or the $D$-term), sometimes dubbed as the ``cosmological constant" of the hadron. 
Since the EMT is obtained from a gauge 
symmetry, it is defined up to a total derivative term whose contributions vanishes upon coupling to the gravitational field. We can choose the total derivative to be $-\partial_\sigma \chi^{\alpha\beta\sigma}$, 
as such the EMT can always be taken to be the free EMT current $t^{\alpha\beta}_\text{f}$. 
Similarly, for spinors, the EMT can be written as the point-like particle EMT plus a total derivative term irrelevant for gravitational coupling. 
Hence we obtain a remarkable conclusion: \textit{the gravitational coupling does not distinguish between elementary and composite particles}. This is the equivalence principle (EP) applied to relativistic quantum systems, and is closely tied to the low-energy theorems of graviton \cite{Weinberg:1964kqu, Weinberg:1964ew, Cho:1976de, Boulware:1974sr}
 which underlines Teryaev's argument for the vanishing of the anomalous gravitomagnetic moment \cite{Teryaev:1999su}. 

Note that Eq.~(\ref{eqn:free_EMT}) differs from the free KG theory ($D=-1$) by some $D$-dependent terms. 
Since the wavepacket is more or less arbitrary, we can consider a static homogeneous wavepacket. In 
this setting, $t^{00}_\text{f} = -DM^2 |\Phi|^2$. The positivity of the energy $t^{00}_\text{f}>0$ immediately leads to $D<0$, which has been speculated for some time as the necessary condition for the stability of matter \cite{Polyakov:2018zvc}. 
Unfortunately, the terms associated with $D$ are non-minimal coupling total derivatives \cite{Parker:2009uva}. 

\section*{Acknowledgements}

The authors wish to thank Y. Chen, X.l. Sheng, S. Pu for valuable discussions.  This work was supported in part by the US Department of Energy (DOE) under Grant Nos. 
DE-FG02-87ER40371 and DE-SC0018223 (SciDAC-4/NUCLEI). Q.W. is supported in part by the National Natural Science Foundation of China (NSFC) under Grants No. 12135011, 11890713, 12047502, and by the Strategic Priority Research Program of the Chinese Academy of Sciences (CAS) under Grant No. XDB34030102. 
Y.L. is supported by the new faculty startup fund of University of Science and Technology of China.


\begin{thebibliography}{99}

\bibitem{Minkowski:1908}
H. Minkowski, Nachrichten von der Gesellschaft der Wissenschaften zu Göttingen, Mathematisch-Physikalische Klasse, 53 (1908);
H. Minkowski, Math. Ann. \textbf{68} (1910) 472

\bibitem{Einstein:1908}
A. Einstein, J. Laub, Annalen Phys. \textbf{331} (1908) 532

\bibitem{Asbury:1967zzb}
J.~G.~Asbury, W.~K.~Bertram, U.~Becker, P.~Joos, M.~Rohde, A.~J.~S.~Smith, S.~Friedlander, C.~Jordan and C.~C.~Ting,
Phys. Rev. Lett. \textbf{18}, 65-70 (1967)
doi:10.1103/PhysRevLett.18.65

\bibitem{Pauli:1981}
Pauli, W., Theory of relativity. Dover Publication (1981); ISBN-13: 978-0486641522

\bibitem{Yennie:1957}
D.R. Yennie, M.M. Lévy, D.G. Ravenhall
Rev. Mod. Phys. \textbf{29} (1957) 144; 

\bibitem{Hofstadter:1958}
R. Hofstadter, F. Bumiller, M.R. Yearian
Rev. Mod. Phys. \textbf{30} (1958) 482.

\bibitem{Ernst:1960zza}
F.~J.~Ernst, R.~G.~Sachs and K.~C.~Wali,
``Electromagnetic form factors of the nucleon,''
Phys. Rev. \textbf{119}, 1105-1114 (1960)
doi:10.1103/PhysRev.119.1105

\bibitem{Sachs:1962zzc}
R.~G.~Sachs,
Phys. Rev. \textbf{126}, 2256-2260 (1962)
doi:10.1103/PhysRev.126.2256

\bibitem{Wong:1998ex}
S.~S.~M.~Wong,
``Introductory nuclear physics,'', 2nd Ed., Wiley-VCH, (2004)

\bibitem{Kelly:2002if}
J.~J.~Kelly,
Phys. Rev. C \textbf{66}, 065203 (2002)
doi:10.1103/PhysRevC.66.065203
[arXiv:hep-ph/0204239 [hep-ph]].

\bibitem{Perdrisat:2006hj}
C.~F.~Perdrisat, V.~Punjabi and M.~Vanderhaeghen,
Prog. Part. Nucl. Phys. \textbf{59}, 694-764 (2007)
doi:10.1016/j.ppnp.2007.05.001
[arXiv:hep-ph/0612014 [hep-ph]].

\bibitem{Suhonen:2007vjh}
J.~Suhonen,
``From Nucleons to Nucleus: Concepts of Microscopic Nuclear Theory,'' Springer, Berlin (2007)
doi:10.1007/978-3-540-48861-3

\bibitem{Pacetti:2014jai}
S.~Pacetti, R.~Baldini Ferroli and E.~Tomasi-Gustafsson,
Phys. Rept. \textbf{550-551}, 1-103 (2015)
doi:10.1016/j.physrep.2014.09.005


\bibitem{Miller:2007uy}
G.~A.~Miller,
Phys. Rev. Lett. \textbf{99}, 112001 (2007)
doi:10.1103/PhysRevLett.99.112001
[arXiv:0705.2409 [nucl-th]].

\bibitem{Miller:2009sg}
G.~A.~Miller,
Phys. Rev. C \textbf{80}, 045210 (2009)
doi:10.1103/PhysRevC.80.045210
[arXiv:0908.1535 [nucl-th]].

\bibitem{Miller:2009qu}
G.~A.~Miller,
Phys. Rev. C \textbf{79}, 055204 (2009)
doi:10.1103/PhysRevC.79.055204
[arXiv:0901.1117 [nucl-th]].

\bibitem{Miller:2018ybm}
G.~A.~Miller,
Phys. Rev. C \textbf{99}, no.3, 035202 (2019)
doi:10.1103/PhysRevC.99.035202
[arXiv:1812.02714 [nucl-th]].


\bibitem{Jaffe:2020ebz}
R.~L.~Jaffe,
Phys. Rev. D \textbf{103}, no.1, 016017 (2021)
doi:10.1103/PhysRevD.103.016017
[arXiv:2010.15887 [hep-ph]].

\bibitem{Freese:2021czn}
A.~Freese and G.~A.~Miller,
Phys. Rev. D \textbf{103}, 094023 (2021)
doi:10.1103/PhysRevD.103.094023
[arXiv:2102.01683 [hep-ph]].

\bibitem{Freese:2021mzg}
A.~Freese and G.~A.~Miller,
Phys. Rev. D \textbf{105}, no.1, 014003 (2022)
doi:10.1103/PhysRevD.105.014003
[arXiv:2108.03301 [hep-ph]].


\bibitem{Dirac:1949cp}
P.~A.~M.~Dirac,
Rev. Mod. Phys. \textbf{21}, 392-399 (1949)
doi:10.1103/RevModPhys.21.392



\bibitem{localization}
There is a long history on the localization of elementary particles, starting as early as 
Pryce [Proc. Roy. Soc. Lond. A \textbf{150}, no.869, 166-172 (1935)]. In short, the localization of relativistic particles in quantum theory is known to be problematic. 
For example, Currie, Jordan and Sudarshan [Rev. Mod. Phys. \textbf{35}, 350-375 (1963)] proved that localized particles must be non-interacting.  Hegerfeldt [Phys. Rev. D \textbf{10}, 3320 (1974)] showed that particle localization is incompatible with causality, which is known to be the case for the famous Newton-Wigner operator \cite{Newton:1949cq}. 
%
See  Refs.~\cite{Wightman:1962sk, Haag:1992hx, Busch:1999, Balachandran:2016bqj} and the references therein for some recent reviews.



\bibitem{Newton:1949cq}
T.~D.~Newton and E.~P.~Wigner,
Rev. Mod. Phys. \textbf{21}, 400-406 (1949)
doi:10.1103/RevModPhys.21.400

\bibitem{Wightman:1962sk}
A.~S.~Wightman,
Rev. Mod. Phys. \textbf{34}, 845-872 (1962)
doi:10.1103/RevModPhys.34.845

\bibitem{Haag:1992hx}
R.~Haag,
1996, Local Quantum Physics: Fields, Particles, Algebras (Springer, Berlin).

\bibitem{Busch:1999}
P. Busch, Phys. A: Math. Gen. 32 (1999) 6535-6546

\bibitem{Balachandran:2016bqj}
A.~P.~Balachandran,
Int. J. Geom. Meth. Mod. Phys. \textbf{14}, no.08, 1740008 (2017)
doi:10.1142/S0219887817400084
[arXiv:1609.01470 [hep-th]].

\bibitem{Burkardt:2000za}
M.~Burkardt,
Phys. Rev. D \textbf{62}, 071503 (2000)
[erratum: Phys. Rev. D \textbf{66}, 119903 (2002)]
doi:10.1103/PhysRevD.62.071503
[arXiv:hep-ph/0005108 [hep-ph]].

\bibitem{Burkardt:2002hr}
M.~Burkardt,
Int. J. Mod. Phys. A \textbf{18}, 173-208 (2003)
doi:10.1142/S0217751X03012370
[arXiv:hep-ph/0207047 [hep-ph]].

\bibitem{Lorce:2020onh}
C.~Lorc\'e,
Phys. Rev. Lett. \textbf{125}, no.23, 232002 (2020)
doi:10.1103/PhysRevLett.125.232002
[arXiv:2007.05318 [hep-ph]].

\bibitem{Epelbaum:2022fjc}
E.~Epelbaum, J.~Gegelia, N.~Lange, U.~G.~Mei\ss{}ner and M.~V.~Polyakov,
[arXiv:2201.02565 [hep-ph]].

\bibitem{Belitsky:2005qn}
A.~V.~Belitsky and A.~V.~Radyushkin,
Phys. Rept. \textbf{418}, 1-387 (2005)
doi:10.1016/j.physrep.2005.06.002
[arXiv:hep-ph/0504030 [hep-ph]].

\bibitem{Belitsky:2003nz}
A.~V.~Belitsky, X.~d.~Ji and F.~Yuan,
Phys. Rev. D \textbf{69}, 074014 (2004)
doi:10.1103/PhysRevD.69.074014
[arXiv:hep-ph/0307383 [hep-ph]].

\bibitem{Ji:2004gf}
X.~Ji,
Ann. Rev. Nucl. Part. Sci. \textbf{54}, 413-450 (2004)
doi:10.1146/annurev.nucl.54.070103.181302


\bibitem{PDG:2020}
P.A. Zyla et al. (Particle Data Group), Prog. Theor. Exp. Phys. \textbf{2020}, 083C01 (2020) and 2021 update.

\bibitem{Polyakov:2018zvc}
M.~V.~Polyakov and P.~Schweitzer,
Int. J. Mod. Phys. A \textbf{33}, no.26, 1830025 (2018)
doi:10.1142/S0217751X18300259
[arXiv:1805.06596 [hep-ph]].


\bibitem{Kharzeev:2021qkd}
D.~E.~Kharzeev,
Phys. Rev. D \textbf{104}, no.5, 054015 (2021)
doi:10.1103/PhysRevD.104.054015
[arXiv:2102.00110 [hep-ph]].

\bibitem{Jackson:1999}
Jackson, J.~D. (1999). Classical Electrodynamics. John Wiley \& Sons, Inc. ISBN 0-471-30932-X.

\bibitem{deGroot:1972}
S.~R. de Groot and L.~G. Suttorp, Foundations of Electrodynamics. 
North-Holland Publishing Company, Amsterdam (1972).

\bibitem{Pohl:2013yb}
R.~Pohl, R.~Gilman, G.~A.~Miller and K.~Pachucki,
Ann. Rev. Nucl. Part. Sci. \textbf{63}, 175-204 (2013)
doi:10.1146/annurev-nucl-102212-170627
[arXiv:1301.0905 [physics.atom-ph]].

\bibitem{Bezginov:2019mdi}
N.~Bezginov, T.~Valdez, M.~Horbatsch, A.~Marsman, A.~C.~Vutha and E.~A.~Hessels,
Science \textbf{365}, no.6457, 1007-1012 (2019)
doi:10.1126/science.aau7807

\bibitem{Xiong:2019umf}
W.~Xiong, A.~Gasparian, H.~Gao, D.~Dutta, M.~Khandaker, N.~Liyanage, E.~Pasyuk, C.~Peng, X.~Bai and L.~Ye, \textit{et al.}
Nature \textbf{575}, no.7781, 147-150 (2019)
doi:10.1038/s41586-019-1721-2

\bibitem{Weinberg:1995mt}
S.~Weinberg,
``The Quantum theory of fields. Vol. 1: Foundations,''   Cambridge University Press (2005).

%
%

\bibitem{Lorce:2021gxs}
C.~Lorc\'e,
Eur. Phys. J. C \textbf{81}, no.5, 413 (2021)
doi:10.1140/epjc/s10052-021-09207-4
[arXiv:2103.10100 [hep-ph]].



\bibitem{Lorce:2017isp}
C.~Lorc\'e,
Phys. Rev. D \textbf{97}, no.1, 016005 (2018)
doi:10.1103/PhysRevD.97.016005
[arXiv:1705.08370 [hep-ph]].

\bibitem{Sheng:2022ssd}
X.~L.~Sheng, Q.~Wang and D.~H.~Rischke,
[arXiv:2202.10160 [nucl-th]].

\bibitem{Polyzou:2012ut}
W.~N.~Polyzou, W.~Gl\"ockle and H.~Witala,
Few Body Syst. \textbf{54}, 1667-1704 (2013)
doi:10.1007/s00601-012-0526-8
[arXiv:1208.5840 [nucl-th]].

\bibitem{Fleming:1974af}
G.~N.~Fleming,
doi:10.1007/978-94-010-2274-3\_22


\bibitem{Panteleeva:2022khw}
J.~Y.~Panteleeva, E.~Epelbaum, J.~Gegelia and U.~G.~Mei\ss{}ner,
[arXiv:2205.15061 [hep-ph]].

\bibitem{Israel:1979wp}
W.~Israel and J.~M.~Stewart,
``Transient relativistic thermodynamics and kinetic theory,''
Annals Phys. \textbf{118}, 341-372 (1979)
doi:10.1016/0003-4916(79)90130-1

\bibitem{Florkowski:2018fap}
W.~Florkowski, A.~Kumar and R.~Ryblewski,
Prog. Part. Nucl. Phys. \textbf{108}, 103709 (2019)
doi:10.1016/j.ppnp.2019.07.001
[arXiv:1811.04409 [nucl-th]].

\bibitem{Sheng:2022wsy}
X.~L.~Sheng, L.~Oliva, Z.~T.~Liang, Q.~Wang and X.~N.~Wang,
[arXiv:2205.15689 [nucl-th]].

\bibitem{Sheng:2022ffb}
X.~L.~Sheng, L.~Oliva, Z.~T.~Liang, Q.~Wang and X.~N.~Wang,
[arXiv:2206.05868 [hep-ph]].



\bibitem{Landau:vol6}
L.~D. Landau, E.~M. Lifshitz, Fluid Mechanics, 2nd Ed. Butterworth-Heinemann (1987),
ISBN 978-0750627672


\bibitem{Eckart:1940te}
C.~Eckart,
Phys. Rev. \textbf{58}, 919-924 (1940)
doi:10.1103/PhysRev.58.919

\bibitem{Selyugin:2009ic}
O.~V.~Selyugin and O.~V.~Teryaev,
Phys. Rev. D \textbf{79}, 033003 (2009)
doi:10.1103/PhysRevD.79.033003
[arXiv:0901.1786 [hep-ph]].

\bibitem{Polyakov:2018zvc}
M.~V.~Polyakov and P.~Schweitzer,
Int. J. Mod. Phys. A \textbf{33}, no.26, 1830025 (2018)
doi:10.1142/S0217751X18300259
[arXiv:1805.06596 [hep-ph]].

\bibitem{Freese:2022fat}
A.~Freese and G.~A.~Miller,
[arXiv:2210.03807 [hep-ph]].

\bibitem{Panteleeva:2022uii}
J.~Y.~Panteleeva, E.~Epelbaum, J.~Gegelia and U.~G.~Mei\ss{}ner,
[arXiv:2211.09596 [hep-ph]].

\bibitem{Alharazin:2022xvp}
H.~Alharazin, B.~D.~Sun, E.~Epelbaum, J.~Gegelia and U.~G.~Mei\ss{}ner,
[arXiv:2212.11505 [hep-ph]].

\bibitem{Rinehimer:2009yv}
J.~A.~Rinehimer and G.~A.~Miller,
Phys. Rev. C \textbf{80}, 015201 (2009)
doi:10.1103/PhysRevC.80.015201
[arXiv:0902.4286 [nucl-th]].


\bibitem{Wiener:1964}
Wiener, N. (1964). Extrapolation, Interpolation, and Smoothing of Stationary Time Series. Cambridge, Mass: MIT Press. ISBN 0-262-73005-7.


\bibitem{Freese:2021jqs}
A.~Freese,
``Noether's theorems and the energy-momentum tensor in quantum gauge theories,''
[arXiv:2112.00047 [hep-th]].

\bibitem{Weinberg:1964kqu}
S.~Weinberg,
Phys. Lett. \textbf{9}, no.4, 357-359 (1964)
doi:10.1016/0031-9163(64)90396-8

\bibitem{Weinberg:1964ew}
S.~Weinberg,
Phys. Rev. \textbf{135}, B1049-B1056 (1964)
doi:10.1103/PhysRev.135.B1049

\bibitem{Boulware:1974sr}
D.~G.~Boulware and S.~Deser,
Annals Phys. \textbf{89}, 193 (1975)
doi:10.1016/0003-4916(75)90302-4

\bibitem{Cho:1976de}
C.~F.~Cho and N.~D.~Hari Dass,
Phys. Rev. D \textbf{14}, 2511 (1976)
doi:10.1103/PhysRevD.14.2511

\bibitem{Teryaev:1999su}
O.~V.~Teryaev,
``Spin structure of nucleon and equivalence principle,''
[arXiv:hep-ph/9904376 [hep-ph]].

\bibitem{Parker:2009uva}
L.~E.~Parker and D.~Toms,
``Quantum Field Theory in Curved Spacetime: quantized field and gravity,'' Cambridge University Press, 2009;
doi:10.1017/CBO9780511813924

\end{thebibliography}
\end{document}